\documentclass[preprintnumbers,superscriptaddress,aps,nofootinbib,perprint,12pt]{revtex4}
\usepackage{amssymb}
\usepackage[centertags]{amsmath}
\usepackage{txfonts}
\usepackage{epsfig}
\usepackage{bm}
\usepackage{color}
\usepackage{graphicx,graphics}
\usepackage{multirow}
\usepackage{float}
\usepackage[pdfstartview=FitH]{hyperref}
\hypersetup{colorlinks=true, citecolor=blue, linkcolor=blue,filecolor=black,urlcolor=blue}
\allowdisplaybreaks[2]

\usepackage{slashed}
\usepackage{ulem}

\usepackage{booktabs}
\usepackage{multirow}

\begin{document}
\title{Energy dependence of hadron polarization in $e^+e^-\to hX$ at high energies}


\author{Kai-bao Chen}
\affiliation{School of Physics \& Key Laboratory of Particle Physics and Particle Irradiation (MOE), Shandong University, Jinan, Shandong 250100, China}

\author{Wei-hua Yang}
\affiliation{School of Physics \& Key Laboratory of Particle Physics and Particle Irradiation (MOE), Shandong University, Jinan, Shandong 250100, China}

\author{Ya-jin Zhou}
\affiliation{School of Physics \& Key Laboratory of Particle Physics and Particle Irradiation (MOE), Shandong University, Jinan, Shandong 250100, China}

\author{Zuo-tang Liang}
\affiliation{School of Physics \& Key Laboratory of Particle Physics and Particle Irradiation (MOE), Shandong University, Jinan, Shandong 250100, China}

\begin{abstract}
The longitudinal polarization of hyperon in $e^+e^-$ annihilation at high energies depends on the longitudinal polarization of the quark 
produced at the $e^+e^-$ annihilation vertex whereas the spin alignment of vector mesons is independent of it. 
They exhibit very different energy dependences. 
We use the longitudinal polarization of Lambda hyperon and the spin alignment of $K^*$ as representative examples 
to present numerical results of energy dependences and demonstrate such distinct differences. 
We present the results at the leading twist with perturbative QCD evolutions of fragmentation functions at the leading order. 
\end{abstract}


\maketitle

\newpage

\section{Introduction} \label{sec:introduction}

The spin dependence of fragmentation functions (FFs) has attracted much attention since it provides 
not only important information on hadronization mechanism but also an important place to study properties of quantum chromodynamics (QCD). 
High energy $e^+e^-$ annihilation is the cleanest place to study FFs. 
Among different aspects of the spin dependence,  vector polarizations of hyperons and tensor polarizations of vector mesons 
are two topics that attracted special attention because both of them can be measured by the angular distributions of the decay products. 
Hyperon polarizations can be determined by the angular distribution of the decay products of the spin self-analyzing parity violating weak decay. 
Different components of the tensor polarization of vector mesons can also be determined by the angular distribution of decay products of the strong decay into two spin zero hadrons.  
Measurements have been carried out e.g. many years ago at LEP for the longitudinal polarization 
of $\Lambda$ hyperon~\cite{Buskulic:1996vb,Ackerstaff:1997nh} 
and for spin alignments of vector mesons~\cite{Ackerstaff:1997kj,Ackerstaff:1997kd,Abreu:1997wd} in the inclusive production process $e^+e^-\to hX$ 
and sizable effects have been observed. 
These data have attracted many phenomenological studies and different approaches have been proposed to describe 
them~\cite{Gustafson:1992iq,Boros:1998kc,Liu:2000fi,Liu:2001yt,Liang:2002ub,Xu:2002hz,
Ma:1998pd,Ma:1999gj,Ma:1999wp,Ma:1999hi,Ma:2000uu,Ma:2000cg,Chi:2013hka,Ellis:2002zv,Anselmino:1997ui,Anselmino:1998jv,Anselmino:1999cg,Xu:2001hz,Xu:2003fq}. 

In the theoretical framework of the QCD parton model, 
hadron polarizations are expressed in terms of different FFs~\cite{Boer:1997mf,Boer:1997qn,Boer:2008fr,Pitonyak:2013dsu,Wei:2013csa,Wei:2014pma,Chen:2016moq}. 
These FFs are defined via quark-quark and/or quark-gluon-quark correlators. 
The results of the complete decomposition of quark-quark correlator as well as those for that of quark-gluon-quark at twist-3 for spin-1 
hadrons has been presented e.g. in~\cite{Chen:2015tca,Chen:2015ora,Chen:2016moq}. 
A general framework for $e^+e^-\to V\pi X$ has been constructed~\cite{Chen:2016moq} and 
QCD parton model results for hadron polarizations in terms of FFs have been presented up to twist-3. 

With these results, we can make predictions on the energy dependence of hadron polarization 
within the theoretical framework of QCD if we know the results at a given energy. 
In fact, from the results presented in~\cite{Boer:1997mf,Boer:1997qn,Boer:2008fr,Pitonyak:2013dsu,Wei:2013csa,Wei:2014pma,Chen:2016moq}, 
we see one distinct feature for hadron polarizations in $e^+e^-$ annihilations at high energies, i.e., at the leading twist, 
polarizations of hadrons are divided into two categories. 
In one of them, the polarization of hadrons depends on the initial longitudinal polarization $P_q$ of the quark (or anti-quark) produced 
at the $e^+e^-$-vertex and is parity violated. 
In the second category, the polarization is independent of $P_q$ and is parity conserved. 
The most well-known example in the inclusive process $e^+e^-\to hX$ is the longitudinal polarization of hyperons such as $\Lambda$, $\Sigma$ and $\Xi$, 
while spin alignments of vector mesons such as $\rho$ and $K^*$ are representatives of the second category.
The longitudinal polarization $P_q$ is a result of weak interaction and is completely determined by the electro-weak process at the parton level.
It takes the maximum for $e^+e^-$ annihilation at the $Z$-pole and changes very fast with energy. 
Hence, we expect that the polarization in the first category has strong energy dependence. 
The energy dependence for hadron polarizations in the second category comes mainly from 
the scale dependence of the corresponding FFs and/or higher twist contributions. 
We expect that they change quite slowly with energy compared with that in the first category. 
We should see very much different behaviors in energy dependence. 

Clearly, the energy dependence provides not only a good place to study the spin dependence of FF 
but also a good place to study QCD evolution of the spin-dependent FF and higher twist contributions.
In view that there are some data available from experiments at LEP~\cite{Buskulic:1996vb,Ackerstaff:1997nh,Ackerstaff:1997kj,Ackerstaff:1997kd,Abreu:1997wd}  
and that new measurements can be carried out in experiments at very much different energies such as BES III and BELLE~\cite{current} 
and possibly at the future facilities planned and/or discussed~\cite{future}, 
it is very interesting to present some numerical results to guide experiment and test models. 
  
In this paper, after a brief summary of hadron polarizations in terms of FFs in $e^+e^-\to hX$, 
we take the longitudinal polarization of $\Lambda$ and the spin alignment of $K^*$ 
as two representative examples for the two categories and calculate the energy dependence. 
We take them as examples because we have data from LEP for both of them. 
We make a simple working parameterization for the corresponding FFs at an initial scale by fitting 
the LEP data~\cite{Buskulic:1996vb,Ackerstaff:1997nh,Ackerstaff:1997kj,Ackerstaff:1997kd,Abreu:1997wd}, 
and evolve them to other energies. 
We present the results numerically that can be used as a rough guide for future experiments. 

The rest of the paper is organized as follows. 
After this introduction, we summarize the results of FFs defined via quark-quark correlator, 
those for hadron polarizations in terms of FFs in $e^+e^-\to hX$ 
and QCD evolution equations for FFs in Sec. II.
In sec. III,  we present a working parameterization of the corresponding FFs, 
show the numerical results of QCD evolution at the leading order and  
present the energy dependence of the two representative examples. 
We make a short summary and an outlook in Sec. IV. 

\section{Hadron polarizations in $e^+e^-\to hX$ in terms of FFs}

High energy $e^+e^-\to hX$ is the best place to study FFs in different connections. 
The results for hadron polarizations expressed in terms of FFs up to twist-3 in leading order in pQCD 
are given in different papers such as~\cite{Boer:1997mf,Boer:1997qn,Boer:2008fr,Pitonyak:2013dsu,Wei:2013csa,Wei:2014pma,Chen:2016moq}.  
Here, we make a short summary of these results and present in particular the formulae that will be used in the numerical estimations. 

\subsection{FFs defined via quark-quark correlator}

The polarization of hadron produced in high energy reaction is described by the spin density matrix. 
For spin-1/2 hadrons,  the polarization is described by a $2\times 2$ spin density matrix that is usually decomposed as $\rho = (1 + \vec S \cdot \vec \sigma)/2$, 
where $\vec\sigma$ is the Pauli matrix, and $\vec S$ is the polarization vector which is represented by the helicity $\lambda$ and 
the transverse polarization vector $S_T^\mu$, i.e.,
\begin{align}
S^\mu = \lambda \frac{p^+}{M} \bar n^\mu + S_T^\mu - \lambda \frac{M}{2p^+}  n^\mu, \label{eq:polVec}
\end{align}
where $n$ and $\bar n$ are the two unit vectors in light-cone coordinates. 
For spin-1 hadrons, the polarization is described by a $3\times 3$ density matrix, which, in the rest frame of the hadron, is usually decomposed as \cite{Bacchetta:2000jk}
\begin{align}
\rho = \frac{1}{3} (\mathbf{1} + \frac{3}{2}S^i \Sigma^i + 3 T^{ij} \Sigma^{ij}), \label{eq:spin1rho}
\end{align}
where $\Sigma^i$ is the spin operator of spin-$1$ particle, 
and $\Sigma^{ij}= \frac{1}{2} (\Sigma^i\Sigma^j + \Sigma^j \Sigma^i) - \frac{2}{3} \mathbf{1} \delta^{ij}$. 
The spin polarization tensor $T^{ij}={\rm Tr}(\rho \Sigma^{ij})$, and is parameterized as
\begin{align}
\mathbf{T}= \frac{1}{2}
\left(
\begin{array}{ccc}
-\frac{2}{3}S_{LL} + S_{TT}^{xx} & S_{TT}^{xy} & S_{LT}^x  \\
S_{TT}^{xy}  & -\frac{2}{3} S_{LL} - S_{TT}^{xx} & S_{LT}^{y} \\
S_{LT}^x & S_{LT}^{y} & \frac{4}{3} S_{LL}
\end{array}
\right).
\label{spintensor}
\end{align}
The tensor polarization part has five independent components that are given by 
a Lorentz scalar $S_{LL}$, a Lorentz vector $S_{LT}^\mu = (0, S_{LT}^x, S_{LT}^y,0)$ 
and a Lorentz tensor $S_{TT}^{\mu\nu}$ that has two nonzero independent components 
$S_{TT}^{xx} = -S_{TT}^{yy}$ and $S_{TT}^{xy} = S_{TT}^{yx}$.

For the fragmentation of the quark (or anti-quark), the FFs are defined via the quark-quark and/or the quark-gluon-quark correlators. 
The quark-quark or quark-gluon-quark correlator can in general be expressed as a sum of a spin-independent part, 
a vector polarization dependent part and a tensor polarization dependent part. 
To describe the production of spin zero hadrons, we need only the spin-independent part. 
For spin-1/2 hadrons, the vector-polarization dependent part is involved, 
and for spin-1 hadrons, the tensor polarization dependent part is also needed.
FFs are obtained by making Lorentz decompositions of the corresponding part 
in terms of 4-momenta and variables describing the polarization. 
Hence, formally, the spin independent part is exactly the same for hadrons with different spins, 
the vector polarization dependent part is also the same for spin-1/2 and spin-1 hadrons.   

The results for the complete decomposition of quark-quark correlator are summarized e.g. in~\cite{Chen:2016moq}. 
At the leading twist, there are totally 18 TMD FFs that are summarized in Table II of \cite{Chen:2016moq}. 
From the table, we see that 5 of these 18 leading twist TMD FFs  
describe fragmentation of unpolarized, 4 of them describe longitudinally polarized and 9 of them describe transversely polarized quark.  
For those describing unpolarized quark fragmentation, we have the well-known $D_1(z,k_\perp)$ describing the number density  
of hadrons produced in the fragmentation and the other 4 describing the induced polarizations. 
Similarly, for FFs of the longitudinally and transversely polarized quark, we have the direct spin transfer $G_{1L}$ and $H_{1T}$  
respectively and others describing the number density and/or ``worm-gear effects''.  

After integrating over the transverse momentum, we obtain the results in the one-dimensional case.  
In this case, we have only five FFs left at the leading twist, i.e., the number density $D_1(z)$, the induced $D_{1LL}(z)$, 
the direct spin transfers in the longitudinally polarized case $G_{1L}(z)$, 
and in the transversely polarized case $H_{1T}(z)$ and $H_{1LT}(z)$.

We emphasize that one-dimensional FFs are needed to describe inclusive processes such as $e^+e^-\to hX$ 
while three-dimensional FFs are needed for semi-inclusive processes such as $e^+e^-\to h_1h_2X$. 
They can be studied in the corresponding processes respectively. 
Also, to study those FFs for unpolarized, transversely polarized or longitudinally polarized quarks, 
one needs to create quarks in the corresponding polarization states and know the polarizations of them before the fragmentation.

\subsection{Quark polarization in $e^+e^-\to q\bar q$}

It is well known that the quark or anti-quark from $e^+e^-\to Z\to q\bar q$ is longitudinally polarized. 
The polarization is given by
\begin{align}
P_q^{Zpole}(\theta) = -\frac{c_1^ec_3^q(1+\cos^2\theta)+2c_3^ec_1^q\cos\theta}{c_1^ec_1^q(1+\cos^2\theta)+2c_3^ec_3^q\cos\theta},
\end{align}
where $\theta$ is the angle between the incident electron and the produced quark, 
$c_1^e = (c_V^{e})^2+(c_A^{e})^2$, $c_3^e = 2 c_V^e c_A^e$, 
$c_V^e$ and $c_A^e$ are defined in the weak interaction current $ \bar \psi \gamma^\mu (c_V^e - c_A^e \gamma^5) \psi$ 
and the superscript denotes that they are for the electron, and similarly for different flavors of quarks.

Although the quark (anti-quark) is not transversely polarized,  their transverse spin components are correlated. 
This is described by the transverse spin correlation function $c_{nn}^q$ defined as  
 \begin{align}
 c_{nn}^q\equiv\frac{|\hat m_{n++}|^2+|\hat m _{n--}|^2-|\hat m_{n+-}|^2-|\hat m_{n-+}|^2}{|\hat m_{n++}|^2+|\hat m_{n--}|^2+|\hat m_{n+-}|^2+|\hat m_{n-+}|^2},
 \end{align}
where $\hat m$ is the scattering amplitude, $+$ or $-$ denotes that the quark or anti-quark is in $s_n=1/2$ or $-1/2$ state. 
If we take $\vec n$ as the normal of the production plane, we obtain
 \begin{align}
c_{nn}^{q,Zpole}(\theta)=\frac{c_1^ec_2^q\sin^2\theta}{c_1^ec_1^q(1+\cos^2\theta)+2c_3^ec_3^q\cos\theta},  \label{eq:cnnq}
 \end{align}
where $c_2^q = (c_V^q)^2-(c_A^q)^2$. 
Define $y=l_2 \cdot p_q/q\cdot p_q\approx (1+\cos\theta)/2$ ($l_1$ and $l_2$ are the 4-momenta of the incident $e^-$ and $e^+$, 
$q=l_1+l_2$ is that of the $Z$-boson, and $p_q$ is that of the produced quark), we can express $P_q$ and $c_{nn}^q$ in terms of $y$, i.e.,   
 \begin{align}
&P_q^{Zpole}(y) = T_{1}^{q} (y) / T_{0}^{q} (y),\\
&c_{nn}^{q,Zpole}(y) = c_1^ec_2^qC(y) / 2T_{0}^{q} (y),\\
& T_0^q (y) =  c_1^e c_1^q  A(y) - c_3^e c_3^q  B(y), \\ 
& T_1^q (y) = - c_1^e c_3^q A(y) + c_3^e c_1^q B(y). 
\end{align}
Here we denote as usual $A(y)=(1-y)^2+y^2\approx(1+\cos^2\theta)/2$, 
$B(y)=1-2y\approx -\cos\theta$, and 
$C(y)=4y(1-y)\approx \sin^2\theta$. 

Experimental studies are often carried out irrespective of $\theta$ or $y$. The obtained results just correspond to the results integrated over $\theta$ or $y$. 
For the polarization and correlation of quark given above, if we integrate over $\theta$ or $y$, we obtain
\begin{align}
&\bar P_q^{Zpole}=-c_3^q/c_1^q,\\
&\bar c_{nn}^{q,Zpole}=c_2^q/2c_1^q.
\end{align}

We see that, the quark is negatively polarized in the longitudinal direction. 
Also $c_2<0$ since $c_V^2$ is smaller than $c_A^2$, so we have a negative $c_{nn}^q$ at the $Z$-pole.  

In general, for $e^+e^-\to q\bar q$, we need to consider contributions from $e^+e^-\to Z\to q\bar q$, 
those from $e^+e^-\to\gamma^*\to q\bar q$ and the interference terms. 
In this case, we have
\begin{align}
&P_q(y) = \Delta w_q(y) / w_q(y),\\
& c_{nn}^q(y) = {2y(1-y)(e_q^2 + \chi c_1^e c_2^q + \chi_{int}^q c_V^e c_V^q)}/w_q(y).
 \end{align}
Here $e_q$ is the electric charge of $q$, and $w_q(y)$ and $\Delta w_q(y)$ are given by
\begin{align}
&w_q(y)= \chi T_{0}^{q} (y) + e_q^2 A (y) + \chi_{int}^q I_0^q (y), \label{eq:wqy}\\ 
&\Delta w_q(y)= \chi T_{1}^{q} (y) + \chi_{int}^q I_1^q (y), \label{eq:dwqy}\\ 
& I_0^q (y) =  c_V^e c_V^q A(y) - c_A^e c_A^q B(y), \\ 
& I_1^q (y) = -c_V^e c_A^q A(y) + c_A^e c_V^q B(y), \\
&\chi={s^2}/{[(s-M_Z^2)^2+\Gamma_Z^2 M_Z^2]\sin^42\theta_W}, \\ 
&\chi_{int}^q = - 2 e_q s (s - M_Z^2) / [(s-M_Z^2)^2 + \Gamma_Z^2 M_Z^2] \sin^2 2\theta_W,
\end{align}
where $M_Z$ and $\Gamma_Z$ are the mass and decay width of $Z$, $\theta_W$ is the Weinberg angle, and $s=q^2=Q^2$. 

\begin{figure}[!ht]
\includegraphics[width=0.45\textwidth]{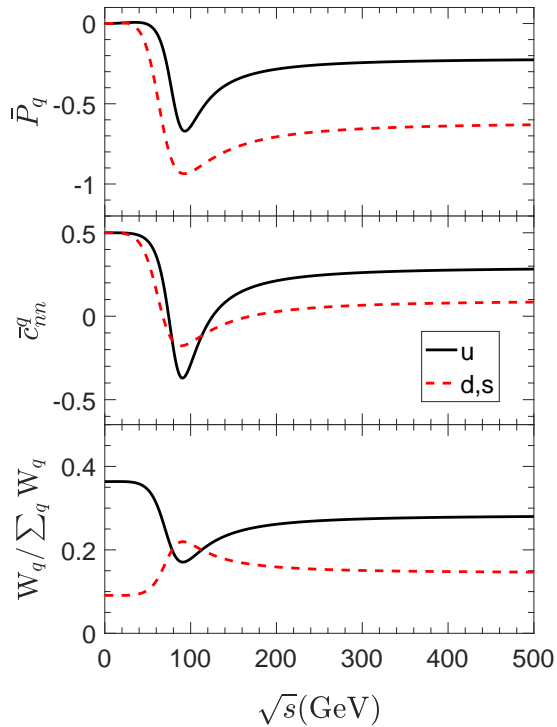}
\caption{Energy dependence of the longitudinal polarization $\bar P_q$, the transverse quark-anti-quark spin correlation $\bar c_{nn}^q$ and the normalized weight $W_q/\sum_qW_q$  
of different flavor $q$ of quark produced in $e^+e^-$ annihilation.} 
\label{fig:Pq}
\end{figure}

After integrating over $y$, we obtain 
\begin{align}
&\bar P_q = \Delta W_q / W_q, \label{eq:Pq}\\
&\bar c_{nn}^q = ({e_q^2 + \chi c_1^e c_2^q + \chi_{int}^q c_V^e c_V^q})/3W_q, \label{eq:cnnq}
\end{align}
where $\Delta W_q$ and $W_q$ are the results of $\Delta w_q(y)$  and $w_q(y)$ after integration over $y$, and they are given by 
\begin{align}
&\Delta W_q= -\frac{2}{3} \bigl( \chi c_1^e c_3^q+ \chi_{int}^q c_V^e c_A^q \bigr), \label{eq:DWq}\\
& W_q=\frac{2}{3} \bigl( e_q^2 + \chi c_1^e c_1^q + \chi_{int}^q c_V^e c_V^q \bigr). \label{eq:Wq}
\end{align}
We see that both $\bar P_q$ and  $\bar c_{nn}^q$ depend on the energy $\sqrt{s}$, 
and behave quite differently in the energy dependence.
For comparison, we plot them in Fig.~\ref{fig:Pq} together with the normalized weight $W_q/\sum_qW_q$.  
We see clearly that, in the energy region $\sqrt{s}\leq M_Z$, as $\sqrt{s}$ decreases, the electromagnetic interaction becomes dominate, 
the longitudinal polarization of quark $\bar P_q$ goes to zero rapidly, 
but the correlation $\bar c_{nn}^q$ goes from negative to positive and reaches the maximum $1/2$ rapidly.
For $\sqrt{s}\ge M_Z$, we have contributions from both weak and electromagnetic interactions, 
and they combine together to give rise to a negative $P_q$ but positive $c_{nn}^q$. 
The correlation between the transverse spin components of the quark and anti-quark is strong and positive. 

From these results we see in particular the following. 
In $e^+e^-$-annihilation at high energies, we have possibilities to 
study FFs of unpolarized, longitudinally polarized as well as transversely polarized quarks. 
First, we can study FFs of unpolarized or longitudinally polarized quarks
by studying singly polarized reactions, i.e., by measuring only the polarization of one hadron in the final state. 
More precisely, we can study one-dimensional FFs of unpolarized or longitudinally polarized quarks in $e^+e^-\to hX$ 
by measuring the corresponding components of polarizations of $h$ in the final state.
By studying the semi-inclusive process $e^+e^-\to h_1h_2X$ and measuring the polarization of $h_1$, 
we can study the corresponding three-dimensional FFs. 
Second, FFs of the transversely polarized quark can also be studied in $e^+e^-$-annihilation at high energies. 
But in this case, we need at least to measure polarizations or other spin dependent asymmetries of two hadrons in the final states 
since the nonzero quantity at the parton level is the transverse spin correlation between the initial quark and anti-quark but not the transverse polarization of the quark or anti-quark.  
In this paper, we start with the simplest case, i.e. $e^+e^-\to hX$ where only one-dimensional FFs for 
the unpolarized or longitudinally polarized quark can be studied.

\subsection{Hadron polarizations at the $Z$-pole}\label{sec:hp_Zpole}

Hadron polarizations in $e^+e^-$-annihilations at high energies are given e.g. in~\cite{Wei:2013csa,Wei:2014pma,Chen:2016moq} in terms of FFs. 
For $e^+e^-\to Z\to VX$ at the leading order in pQCD and up to twist-3, for the longitudinal components, we obtain
\begin{align}
&\langle \lambda\rangle(z,y) 
= \frac{2}{2S+1} \frac{\sum_q P_q(y)T_0^q(y) G_{1L}(z)}{\sum_q T_0^q(y) D_1(z)}, \label{eq:zLambda}\\
&\langle S_{LL}\rangle(z,y)
= \frac{3}{2(2S+1)} \frac{\sum_q T_0^q(y)D_{1LL}(z)}{\sum_q T_0^q(y)D_1(z)}. \label{eq:zSll}
\end{align}
Here, for brevity and clarity, we omit the superscript $q\to V$ in the fragmentation functions, e.g., $D_1(z) = D_1^{q \to V}(z)$; 
and $S$ is the spin of hadron $h$. The factor $(2S+1)$ appears here because, in the conventions used in~\cite{Chen:2016moq} 
in defining FFs via quark-quark correlator and/or quark-gluon-quark correlator, $D_1(z)$ is the number density for the produced $h$ 
averaging over rather than summing over the spin of $h$. 
We write this factor explicitly so that the corresponding expressions eventually take the same form for spin-1/2 as well as spin-1 hadrons. 
For the transverse components, we have, 
\begin{align}
&\langle S_T^{x}\rangle(z,y) =-\frac{8MD(y)}{(2S+1)zQ} \frac{\sum_q T_3^q(y) G_T(z)}{\sum_q T_0^q(y) D_1(z)}, \label{eq:zStx}\\
&\langle S_T^{y}\rangle(z,y) =\frac{8MD(y)}{(2S+1)zQ} \frac{\sum_q T_2^q(y) D_T(z)}{\sum_q T_0^q(y) D_1(z)}, \label{eq:zSty}\\
&\langle S_{LT}^{x}\rangle(z,y) =-\frac{8MD(y)}{(2S+1)zQ} \frac{\sum_q T_2^q(y) D_{LT}(z)}{\sum_q T_0^q(y) D_1(z)}, \label{eq:zSltx}\\
&\langle S_{LT}^{y}\rangle(z,y) =\frac{8MD(y)}{(2S+1)zQ} \frac{\sum_q T_3^q(y) G_{LT}(z)}{\sum_q T_0^q(y) D_1(z)}, \label{eq:zSlty}
\end{align}
where $D(y) = \sqrt{y(1-y)}$, and we also define,
\begin{align}
& T_2^q (y) = - c_3^e c_3^q  + c_1^e c_1^q  B(y), \\ 
& T_3^q (y) =   c_3^e c_1^q  - c_1^e c_3^q  B(y).
\end{align}
We recall that $\langle S_T^{y}\rangle$ is $P$-even and naive $T$-odd, $\langle S_T^{x}\rangle$ is $P$-odd and naive $T$-even, 
and $\langle S_{LT}^{y}\rangle$ is $P$-odd and naive $T$-odd. 
We emphasize that formally vector polarization components such as $\langle \lambda \rangle$, $\langle S_T^{x}\rangle$  
and $\langle S_{T}^{y}\rangle$ have exactly the same expressions in terms of FFs for spin-1/2 hadrons or vector mesons. 
This means that Eqs.~(\ref{eq:zLambda}) and (\ref{eq:zStx}-\ref{eq:zSty}) are the same for hyperons and for vector mesons. 
They are just given by the corresponding FFs for specified hadrons.

The spin alignment of the vector meson is measured by the $00$-component $\rho_{00}$ of the spin density matrix $\rho$ in the helicity base.
It is directly related to $\langle S_{LL}\rangle$ by $\rho_{00}=(1-2\langle S_{LL}\rangle)/{3}$,  
which means 
\begin{align}
\rho_{00}(z,y) = \frac{1}{3} - \frac{1}{3} \frac{\sum_q T_0^q(y) D_{1LL}(z)}{\sum_q T_0^q(y) D_1(z)}.
\end{align}

We consider the case of integrated over $\theta$ or $y$, and we have
\begin{align}
&\bar \lambda (z) 
= -\frac{2}{2S+1} \frac{\sum_q c_3^q G_{1L}(z) }{\sum_q c_1^q D_1(z) }, \label{eq:AzLambda}\\
&\bar\rho_{00}(z) = \frac{1}{3} - \frac{1}{3} \frac{\sum_q c_1^q D_{1LL}(z)}{\sum_q c_1^q D_1(z)}, \\
&\bar S_T^{x}(z) =-\frac{3\pi M}{2(2S+1)zQ} \frac{\sum_q c_3^e c_1^q G_T(z) }{\sum_q c_1^e c_1^q D_1(z) }, \label{eq:AzStx}\\
&\bar S_T^{y}(z) =-\frac{3\pi M}{2(2S+1)zQ} \frac{\sum_q c_3^e c_3^q D_T(z) }{\sum_q c_1^e c_1^q D_1(z) }, \label{eq:AzSty}\\
&\bar S_{LT}^{x}(z) =\frac{3\pi M}{2(2S+1)zQ} \frac{\sum_q c_3^e c_3^q D_{LT}(z)}{\sum_q c_1^e c_1^q D_1(z) }, \label{eq:AzSltx}\\
&\bar S_{LT}^{y}(z) =\frac{3\pi M}{2(2S+1)zQ} \frac{\sum_q c_3^e c_1^q G_{LT}(z) }{\sum_q c_1^e c_1^q D_1(z) }. \label{eq:AzSlty} 
\end{align}

We see that at the leading twist, we have only two non-zero components, i.e. the longitudinal polarization $P_{Lh}=\langle\lambda\rangle$ and 
$\rho_{00}=(1-2\langle S_{LL}\rangle)/3$. 
The transverse polarization exists at twist-3, i.e., it is power suppressed. 
We also note that there is no twist-3 contribution to $\langle\lambda\rangle$ or $\langle S_{LL}\rangle$.
The higher twist corrections to these two components come only from twist-4 or even higher twists~\cite{Wei:2013csa}. 

\subsection{Hadron polarizations at different energies}\label{sec:hp_energy}

At different energies, we need to consider contributions from $e^+e^-\to Z\to VX$, those from $e^+e^-\to \gamma^* \to VX$ 
and those from the interference terms.   
For the longitudinal components, we have 
\begin{align}
&\langle \lambda\rangle (z,y)
= \frac{2}{2S+1} \frac{\sum_q P_q(y) w_q(y) G_{1L}(z)}{\sum_q w_q(y) D_1(z)}, \label{eq:Lambda}\\
&\langle S_{LL}\rangle (z,y) = \frac{3}{2(2S+1)} \frac{\sum_q w_q(y)D_{1LL}(z)}{\sum_q w_q(y)D_1(z)}, \label{eq:Sll}
\end{align}
and for the transverse components
\begin{align}
&\langle S_T^{x}\rangle (z,y) =-\frac{8M D(y)}{(2S+1)zQ} \frac{\sum_q \Delta_x w_q(y) G_T(z)}{\sum_q w_q(y) D_1(z)}, \label{eq:Stx}\\
&\langle S_T^{y}\rangle (z,y) =\frac{8M D(y)}{(2S+1)zQ} \frac{\sum_q \Delta_y w_q(y) D_T(z)}{\sum_q w_q(y) D_1(z)}, \label{eq:Sty}\\
&\langle S_{LT}^{x}\rangle (z,y) =-\frac{8M D(y)}{(2S+1)zQ} \frac{\sum_q \Delta_y w_q(y) D_{LT}(z)}{\sum_q w_q(y) D_1(z)}, \label{eq:Sltx}\\
&\langle S_{LT}^{y}\rangle (z,y) =\frac{8M D(y)}{(2S+1)zQ} \frac{\sum_q \Delta_x w_q(y) G_{LT}(z)}{\sum_q w_q(y) D_1(z)}, \label{eq:Slty}
\end{align}
where $w_q(y)$ is given by Eq.~(\ref{eq:wqy}), and $\Delta_x w_q(y)$ and $\Delta_y w_q(y)$ are given by
\begin{align}
&\Delta_x w_q(y) = \chi T_3^q (y) + \chi_{int}^q  I_3^q (y), \\
&\Delta_y w_q(y) = \chi T_2^q (y) + \chi_{int}^q I_2^q (y), \\
& I_2^q (y) = - c_A^e c_A^q  + c_V^e c_V^q  B(y), \\
& I_3^q (y) =   c_A^e c_V^q - c_V^e c_A^q B(y).
\end{align}
After integrating over $y$, we obtain
\begin{align}
& \bar\lambda(z) = \frac{2}{(2S+1)} \frac{\sum_q \bar P_q W_q G_{1L}(z)}{\sum_q W_q D_1(z)},\label{eq:ALambda}\\
&\bar \rho_{00}(z) = \frac{1}{3} - \frac{1}{3} \frac{\sum_q W_q D_{1LL}^{q\to h}(z)}{\sum_q W_q D_1^{q\to h}(z)},\label{eq:Arho00}\\ 
& \bar S_T^x (z) = -\frac{8M}{(2S+1)zQ} \frac{\sum_q \Delta_x W_q G_T(z)}{\sum_q W_q D_1(z)},\label{eq:AStx}\\
& \bar S_T^y (z) = \frac{8M}{(2S+1)zQ} \frac{\sum_q \Delta_y W_q D_T(z)}{\sum_q W_q D_1(z)}, \label{eq:ASty}\\
&\bar S_{LT}^{x} (z) =-\frac{8M}{(2S+1)zQ} \frac{\sum_q \Delta_y W_q D_{LT}(z)}{\sum_q W_q D_1(z)}, \label{eq:ASltx}\\
&\bar S_{LT}^{y} (z) =\frac{8M}{(2S+1)zQ} \frac{\sum_q \Delta_x W_q G_{LT}(z)}{\sum_q W_q D_1(z)}, \label{eq:ASlty}
\end{align}
where $\Delta_x W_q = {\pi}(\chi c_3^e c_1^q  + \chi_{int}^q c_A^e c_V^q )/8$, and 
$\Delta_y W_q = -{\pi}(\chi c_3^e c_3^q  + \chi_{int}^q c_A^e c_A^q )/8$. 
We see again that there exist twist-3 transverse polarizations that can be used to study 
higher twist effects in particular the corresponding higher twist FFs. 
However, we should also note that at lower energies where electromagnetic interactions dominate, 
such twist-3 contributions are non-zero only at a given $y$ but vanish after the integration over $y$ or $\theta$ in the entire region. 
This is consistent with the data available~\cite{Althoff:1984iz}. 
One can however study such effects by measuring transverse polarizations integrated in a given region of $\theta$ or $y$ 
such as in the forward or backward hemisphere.

\subsection{QCD evolution equations for $G_{1L}$ and $D_{1LL}$}

QCD evolutions for leading twist FFs have been well established and are determined by corresponding 
DGLAP equations~\cite{Dokshitzer:1977sg,Gribov:1972ri,Gribov:1972rt,Altarelli:1977zs} 
with time-like splitting functions~\cite{Owens:1978qz,Georgi:1977mg,Uematsu:1978yw}.
We just give the equations that will be used in our numerical estimations in the following. 
The evolution of the spin transfer $G_{1L}$ is given by DGLAP in the longitudinally polarized case 
while that for the $S_{LL}$-dependent FF $D_{1LL}$ is the same as that for unpolarized FF $D_1$. 
They are given by 
\begin{align}
\frac{\partial}{\partial \ln Q^2} G_{1L}^{i\to h} (z,Q^2)
&=\frac{\alpha_s(Q^2)}{2\pi} \sum_j \int_z^1\frac{dy}{y}G_{1L}^{j\to h}(\frac{z}{y},Q^2) \Delta P_{ji}(y,\alpha_s),\label{eq:DGLAPG1L}\\
\frac{\partial}{\partial \ln Q^2} D_{1LL}^{i\to h} (z,Q^2)
&=\frac{\alpha_s(Q^2)}{2\pi} \sum_j \int_z^1\frac{dy}{y}D_{1LL}^{j\to h}(\frac{z}{y},Q^2) P_{ji}(y,\alpha_s),\label{eq:DGLAPD1LL}
\end{align}
where $i$ or $j$ denotes different types of partons such as different flavors of quarks, anti-quarks and gluon. 
At the leading order (LO) in pQCD, the polarized splitting functions are given by~\cite{Ravindran:1996ri,Ravindran:1996jd} 
\begin{align}
\Delta P_{qq}(y)=&C_F\left[\frac{1+y^2}{(1-y)_+}+\frac{3}{2}\delta(1-y)\right], \\
\Delta P_{gq}(y)=&C_F\frac{1-(1-y)^2}{y},\\
\Delta P_{qg}(y)=&[y^2-(1-y)^2]/2,\\
\Delta P_{gg}(y)=&N_c \left[ (1+y^4)\Bigl(\frac{1}{y}+\frac{1}{(1-y)_+}\Bigr)- \frac{(1-y)^3}{y}\right] 
+\frac{11N_c-2N_f}{6}\delta(1-y), 
\end{align}
where $N_c=3$ and $C_F=(N_c^2-1)/2N_c$.  
The unpolarized splitting functions are given by, 
\begin{align}
P_{qq}(y)=&\Delta P_{qq}(y),\\ 
P_{gq}(y)=&C_F\frac{1+(1-y)^2}{y},\\
P_{qg}(y)=&[y^2+(1-y)^2]/2,\\
P_{gg}(y)=&N_c \left[ \frac{2y}{(1-y)_+} - 2(y^2-y-\frac{1}{y}+1) \right] 
+\frac{11N_c-2N_f}{6}\delta(1-y).
\end{align}

The next-to-leading (NLO) results for these splitting functions have also been obtained  
and a global fit for the spin-dependent FF $G_{1L}^{q\to\Lambda}(z,Q)$ have been given in~\cite{deFlorian:1997zj}. 
However the data available are still too far from enough to make such detailed analysis for other hadrons. 
Even for $G_{1L}^{q\to\Lambda}(z,Q)$, we still far away from a reliable parameterization 
of different contributions~ \cite{deFlorian:1997zj}.  
The purpose of our studies in this paper is not to make a global fit for polarized FFs but to   
demonstrate the two distinctly different behaviors in the energy dependence of hadron polarization in $e^+e^-$-annihilation. 
We therefore limit ourselves to the next-to-leading order in pQCD 
where only leading order splitting functions given above are used. 

\section{Numerical results for $P_{L\Lambda}$ and $\bar \rho_{00}^{K^*}$}

As has already been emphasized in Sec.~II, from the results given by Eqs.~(\ref{eq:zLambda}-\ref{eq:zSlty}) 
and (\ref{eq:Lambda}-\ref{eq:Slty}), we see clearly that 
at the leading twist there exist only two components of the polarization, $P_{Lh}=\bar\lambda$ and $\bar S_{LL}$ or $\bar\rho_{00}$, 
and that there is a distinct difference between them: 
The former depends on the initial longitudinal polarization $\bar P_q$ of the quark 
and describes the longitudinal spin transfer in the fragmentation of the quark.  
It is parity violating in $e^+e^-\to hX$ and is caused by the weak channel $e^+e^-\to Z\to hX$ and its interference to electromagnetic channel $e^+e^-\to \gamma^* \to hX$. 
The latter is independent of $\bar P_q$ and is an induced polarization in the quark fragmentation. 
It is parity conserved and exists even in the fragmentation of the unpolarized quark. 
Clearly such distinct differences should manifest themselves in different high energy reactions. 
One of the consequences in $e^+e^-$-annihilation is the different behavior in the energy dependence. 

As can be seen from Eq.~(\ref{eq:ALambda}), the energy dependence of $P_{Lh}$ comes from three factors: 
the quark polarization $P_q$, the relative weight $W_q$ of different flavors and the scale dependence of FFs. 
The first two factors are determined by the electro-weak interactions and can easily be calculated. 
From the results (see e.g. Fig.~\ref{fig:Pq}), we see clearly that the energy dependence of $\bar P_q$ 
is very strong in particular in the region for $\sqrt{s}$ a bit larger than $M_Z$ down to few GeV, say, $5<\sqrt{s}<200$GeV. 
That of the normalized $W_q$ in this energy region is also quite obvious but quite smother than that of $\bar P_q$. 
Furthermore, the influence of $W_q$ on the polarization of hadron can only be transferred via the flavor dependence of the corresponding FFs 
and can be weakened in kinematic regions where the flavor dependence of FF is not strong.  
The scale dependence of FF is determined by QCD evolution and is perhaps the smoothest among the three factors~\cite{deFlorian:1997zj} in the above-mentioned energy region. 
Hence, without detailed calculations, we can already expect that  the longitudinal polarization of the hyperon in $e^+e^-$-annihilation 
changes fast with energy and the behavior is dominated by that of $\bar P_q$. 
In contrast, for $\bar S_{LL}$, the energy dependence comes only from that of $W_q$ and FFs. 
There should be a much smother energy dependence for $S_{LL}$ in the energy region mentioned above. 
This is a clear prediction based on the general features of the QCD quark-parton picture 
and is independent of the detailed behaviors of FFs that can be tested by experiments. 

Currently, our knowledge on the precise forms of FFs are still very limited due to limitations of data available in particular in the polarized case. 
It is quite fortunate that there are some data available from LEP~\cite{Buskulic:1996vb,Ackerstaff:1997nh,Ackerstaff:1997kj,Ackerstaff:1997kd,Abreu:1997wd} 
on the longitudinal polarization $P_{L\Lambda}$ of $\Lambda$ and the spin alignment $\bar \rho_{00}^{K^*}$ of $K^*$. 
Although they are still very far from sufficient for a detailed analysis, we can use them to initialize such a study to demonstrate
the essential features and guide future experimental studies.

\subsection{Parameterizations and QCD evolutions of $G_{1L}$ and $D_{1LL}$}\label{FFpar}

Because of decay contributions, polarization of $\Lambda$ hyperon is much more involved than other hyperons and/or vector mesons. 
In general,  the leading twist FF for a quark to a baryon, $q\to B_i$, can be written as
the sum of a direct fragmentation and a decay contribution part, i.e., 
\begin{align}
D_1^{q \to B_i}(z) = &D_{1dir}^{q\to B_i}(z)+\sum_{j\ne i} R_D^{j i} 
\int dz'  K_D^{ji}(z,z') D_1^{q \to B_j}(z'),\\
G_{1L}^{q \to B_i}(z) =& G_{1L,dir}^{q\to B_i}(z) +\sum_{j\ne i}R_D^{ji} 
\int dz'  K_D^{ji}(z,z') t_D^{ji}(z,z') G_{1L}^{q \to B_j}(z'),
\end{align}
where $R_D^{ji}$ is the decay branch ratio of $B_j \to B_i+X$, 
$K_D^{ji}(z,z')$ is the probability for a $B_j$ with $z'$ to decay into a $B_i$ with $z$, 
and $t_D^{ji}(z,z')$ is the spin transfer factor in the decay process.

Numerical results show that, for the $\Lambda$ hyperon, the contributions from $\Sigma^0$ and $\Xi^{0,-}$ are sizable~\cite{Boros:1998kc,Liu:2000fi}.
However, since there is no suitable data for $\Sigma^0$ or $\Xi$ polarization in $e^+e^-$ available yet, 
it is impossible to make such a detailed analysis. 
On the other hand, the energy dependences of the hadron polarizations that we will study in this paper come mainly from the QCD evolution of FFs 
and the energy dependence of the polarization of the quark produced at the $e^+e^-$-annihilation vertex. 
We would expect that the influences from the decay contributions on the energy dependence are not very large. 
Hence, in this paper, as a rough estimation, 
we simply parameterize the final $D_1^{q \to \Lambda}(z)$ and $G_{1L}^{q \to \Lambda}(z)$, 
and evolve them to other energies using DGLAP equations given by Eq.~(\ref{eq:DGLAPG1L}). 

Currently, for the unpolarized FF $D_1(z)$, there exist already a number of parameterizations in literature 
for the production of hadrons such as pion, Kaon, proton and $\Lambda$~\cite{Agashe:2014kda}. 
We can just take the most recent parameterizations AKK08 given in~\cite{Albino:2008fy} for $\Lambda$.

For the polarized FF $G_{1L}^{q \to \Lambda}(z)$, 
a global fit and detailed analysis have already been given in 1998 in~\cite{deFlorian:1997zj} 
by de Florian, Stratmann and Vogelsang (DSV98) to the NLO in QCD evolution. 
However, as have already been pointed out in~\cite{deFlorian:1997zj}, the data available are far from sufficient to fix all different contributions. 
They had to make some assumptions such as that the heavy-flavor contributions are neglected, that the $u$ and $d$ fragmentations are taken as equal, 
and that the polarized unfavored and gluon FFs are taken as zero at the initial scale etc in order to carry out the calculations. 
They presented also the results at different scenarios. 
The results already show explicitly that, compared to the drastic change of $\bar P_q$ with energy shown in Fig.~\ref{fig:Pq}, 
the scale dependence of FF is smooth and the difference between LO and NLO results is not very large.  

There is not much progress on parameterizations of polarized FFs since DSV98~\cite{deFlorian:1997zj} besides some improvements for unpolarized FFs. 
There are however many phenomenological studies~\cite{Gustafson:1992iq,Boros:1998kc,Liu:2000fi,Liu:2001yt,Liang:2002ub,Xu:2002hz,
Ma:1998pd,Ma:1999gj,Ma:1999wp,Ma:1999hi,Ma:2000uu,Ma:2000cg,Chi:2013hka,Ellis:2002zv,Zuotang:2002ub,Dong:2005ea,Erkol:2009ek}  
using different models on hyperons polarization in different high energy reactions.  
We note in particular a series of analysis with the aids of the Monte-Carlo event generator such as 
PYTHIA based on Lund string fragmentation model~\cite{Sjostrand:1993yb,Andersson:1983ia} 
and another series\cite{Ma:1998pd,Ma:1999gj,Ma:1999wp,Ma:1999hi,Ma:2000uu,Ma:2000cg,Chi:2013hka}   
based on the Gribov relation~\cite{Gribov:1971zn} between PDFs and FFs. 
With a Monte-Carlo event generator, one can analyze in detail the influence of different contributions.  
From these different phenomenological studies~\cite{Gustafson:1992iq,Boros:1998kc,Liu:2000fi,Liu:2001yt,Liang:2002ub,Xu:2002hz,
Ma:1998pd,Ma:1999gj,Ma:1999wp,Ma:1999hi,Ma:2000uu,Ma:2000cg,Chi:2013hka,Ellis:2002zv,Zuotang:2002ub,Dong:2005ea,Erkol:2009ek}, 
we see that although there are distinct differences in different models, some features are common. 
Consistent with the data available~\cite{Buskulic:1996vb,Ackerstaff:1997nh,Ackerstaff:1997kj,Ackerstaff:1997kd,Abreu:1997wd}, 
all models seem to suggest that polarization dependent FFs are important only in the large $z$ region. 
In the language of the Feynman-Field type of recursive cascade fragmentation picture~\cite{Field:1976ve,Feynman:1977yr,Field:1977fa,Feynman:1978dt}, 
polarized FFs are dominated by the contributions of the ``first rank'' hadrons, i.e. those containing the fragmenting quarks and/or anti-quarks. 
It implies also that the main features are determined by the ``favored'' fragmentations. 
The contributions from the ``unfavored'' and gluon fragmentations are from ``higher rank'' hadrons. 
They are in general small and have negligible dependence on the flavor of the initial quark (anti-quark) or gluon. 
This is also consistent with the assumptions made by DSV in~\cite{deFlorian:1997zj}.   

\begin{figure}[!ht]
\includegraphics[width=0.5\textwidth]{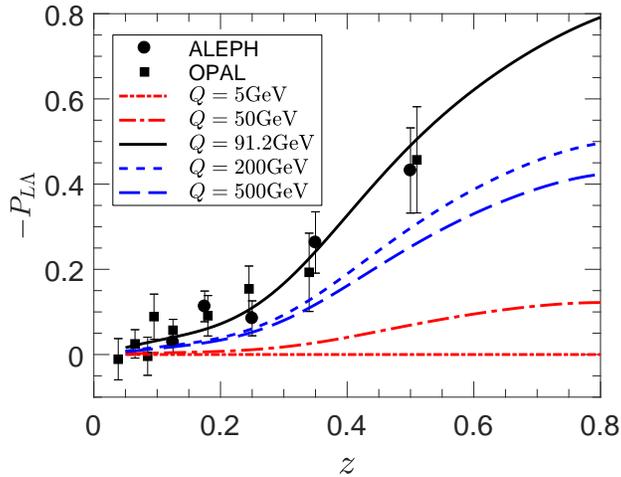}
\caption{(color online) Longitudinal polarization of $\Lambda$ in $e^+e^-\to\Lambda X$ at high energies. 
The LEP data are taken from~\cite{Buskulic:1996vb,Ackerstaff:1997nh}. 
The solid line is the fit described in the text while those at other energies are calculated results 
using DGLAP for FFs and energy dependence of $P_q$. } 
\label{fig:fit_Lambda}
\end{figure}

\begin{figure}[!ht]
\includegraphics[width=0.5\textwidth]{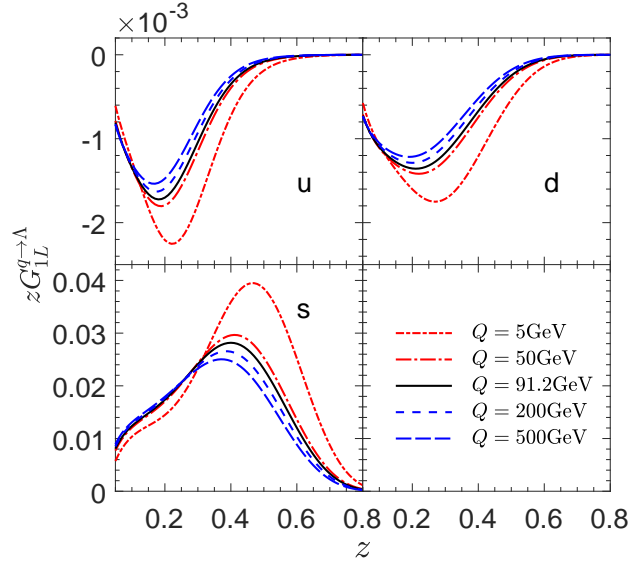}
\caption{(color online) The longitudinal spin transfer fragmentation function $G_{1L}(z,Q^2)$ for $q \to \Lambda$ as a function of $z$ for different flavors of $q$ at different values of $Q$. 
The solid lines are obtained by fitting data for $P_{L\Lambda}$ at $Q=M_Z$, and the others are obtained using DGLAP with leading order splitting functions described in the text. } 
\label{fig:G1Lq2lambda}
\end{figure}

Since the purpose of the calculations in this section is to demonstrate the main feature of the energy dependence of hyperon polarization, 
we do not intend to make a best fit for the data available. 
Instead, we would like to pick up the most influential parts to demonstrate the main features expected. 
In view that the differences between LO QCD and NLO are not very large~\cite{deFlorian:1997zj} and the accuracy that we can reach at this stage is not very high,  
we choose to do only LO QCD evolution. 
We take a parametrization in the same form as DSV98~\cite{deFlorian:1997zj} in the second scenario 
with the same assumptions and/or approximations in connection with the heavy flavor contributions, $u$ and $d$ flavor dependence, unfavored and gluon fragmentation at the initial scale.  
We will use the most recent parameterization for the corresponding unpolarized FFs and re-adjust the parameters to get a better fit to the LEP data~\cite{Buskulic:1996vb,Ackerstaff:1997nh}.  
More precisely, for the $s$-quark fragmentation, we take
\begin{align}
& G_{1L}^{s\to\Lambda}(z) = z^a D_{1}^{s\to\Lambda}(z), 
\end{align}
while for $u$ and $d$-quark, we take 
\begin{align}
& G_{1L}^{q\to\Lambda}(z) = N z^a D_{1}^{q\to\Lambda}(z), 
\end{align}
where $q=u$ or $d$, and limit the parameters $a>0$ and $|N| \leqslant 1$ so that the positivity bounds are satisfied~\cite{deFlorian:1997zj}. 
We further limit $N$ to be negative and small to consist with the expectation from the Gribov relation~\cite{Gribov:1971zn} and polarized PDFs~\cite{Jaffe:1996wp}.  
We choose an initial scale at $Q=5$GeV evolves the FFs to larger $Q$'s and fix the parameters by the LEP data for $\Lambda$ polarization\cite{Buskulic:1996vb,Ackerstaff:1997nh}. 
In this way, we fix the parameters as $a = 0.5$ and $N = -0.1$. 
The result of the fit for $\Lambda$ polarization is shown by the solid line in Fig. \ref{fig:fit_Lambda}. 
The obtained $G_{1L}^{q\to\Lambda}(z,Q^2)$ at different $Q$ for  $q=u$, $d$ and $s$ are shown in Fig.~\ref{fig:G1Lq2lambda}.
We see that in general $G_{1L}^{s\to\Lambda}(z,Q^2)$ is positive and much larger than $G_{1L}^{u\to\Lambda}(z,Q^2)$ or $G_{1L}^{d\to\Lambda}(z,Q^2)$. 
The small difference between $G_{1L}^{u\to\Lambda}(z,Q^2)$ and $G_{1L}^{d\to\Lambda}(z,Q^2)$ comes from that 
between $D_{1}^{u\to\Lambda}(z,Q^2)$ and $D_{1}^{d\to\Lambda}(z,Q^2)$ from AKK08~\cite{Albino:2008fy}.


For vector mesons such as $K^*$, the decay contributions are negligible.  
So we need to simply  parameterize and evolve the corresponding $D_1$ and $D_{1LL}$ to obtain $\bar S_{LL}$ at different energies. 
However, our knowledge about parameterizations of the corresponding FFs is even more limited than that for $\Lambda$. 
There is even no parameterization of FF in the unpolarized case available yet.  
As a start, we make a simple parameterization by using those for $K^{\pm}$ from AKK08~\cite{Albino:2008fy} 
and by parameterizing the data for the ratio of $K^*$ to $K$ as given in~\cite{Shlyapnikov:2001jf}. 
The $z$-dependence for the ratio is taken as linear, i.e., $D_1^{K^*}(z)/3D_1^{K^+}(z) = 0.2z + 0.1$, 
and is assumed to be the same for different flavors.  

\begin{figure}[!ht]
\includegraphics[width=0.5\textwidth]{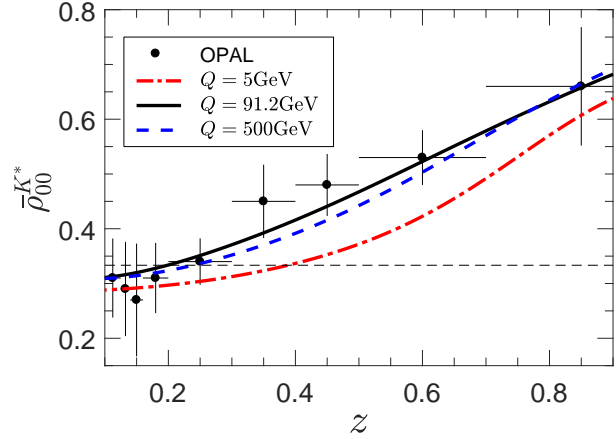}
\caption{(color online) Spin alignment of $K^*$ as a function of $z$. 
The solid line is the fit described in the text while those at other energies are calculated results using DGLAP for FFs. 
The data points are from OPAL at LEP and are taken from~\cite{Ackerstaff:1997kj}.} 
\label{fig:fit_SpinAlignment}
\end{figure}

\begin{figure}[!ht]
\includegraphics[width=0.5\textwidth]{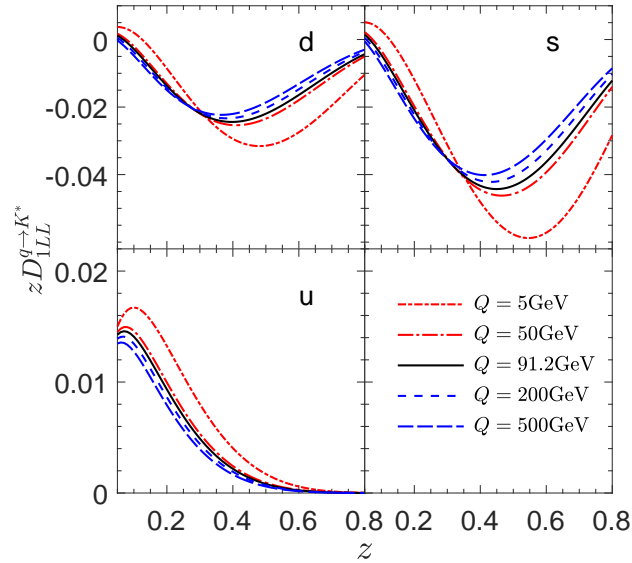}
\caption{(color online) The $S_{LL}$ dependent fragmentation function $D_{1LL}(z,Q^2)$ for $q \to K^*$ as a function of $z$ for different flavors of $q$ at different values of $Q$. 
The solid lines are obtained by fitting data for $\rho_{00}^{K^*}$ at $Q=M_Z$, the others are obtained using DGLAP with leading order splitting functions described in the text.} 
\label{fig:D1LLq2Kstar}
\end{figure}

For the $S_{LL}$-dependent FFs, $D_{1LL}^{q\to K^*}(z)$, we carry out the calculations at the same level as that given above for $G_{1L}^{q\to \Lambda}(z)$. 
Inspired by the almost linear $z$-dependence of data of $\rho_{00}$~\cite{Ackerstaff:1997kj}, 
we parameterize $D_{1LL}^{q\to K^*}(z)$ as 
\begin{align}
& D_{1LL}^{q\to K^*}(z) = c_1 D_{1}^{q\to K^*}(z), 
\end{align}
for un-favored fragmentations and
\begin{align}
& D_{1LL}^{q\to K^*}(z) = (c_1+c_2 z) D_{1}^{q\to K^*}(z), 
\end{align}
for favored fragmentations. 
We limit $-3/2<c_1<3$ and $-3/2<c_1+c_2 z<3$ for $0<z<1$ to satisfy the positivity bound. 
By fitting the available $z$-dependence data on the spin alignment of $K^*$ from OPAL~\cite{Ackerstaff:1997kj},
we fix the parameters as $c_1 = 0.15$ and $c_2 = -1.2$ at $Q$=5GeV. 
The fitted curve is presented in Fig.~\ref{fig:fit_SpinAlignment}.
The obtained $D_{1LL}^{q\to K^{*0}}(z,Q^2)$ at different $Q$ for $q=u$, $d$ or $s$ is given by the solid line in Fig.~\ref{fig:D1LLq2Kstar}.

Because the data (see Fig.~\ref{fig:fit_SpinAlignment}) for $\rho_{00}^{K^*}$ are larger than $1/3$ in the large $z$ region, 
the $S_{LL}$-dependent FF $D_{1LL}(z,Q^2)$ should be negative in the corresponding $z$ region.  
At small $z$, $\rho_{00}$ is smaller than $1/3$, which implies a positive $D_{1LL}(z,Q^2)$. 
These features are shown clearly in Fig.~\ref{fig:D1LLq2Kstar}, 
where we can see that for favored fragmentations, $D_{1LL}(z,Q^2)$ are negative at larger $z$ 
while those for unfavored fragmentations also play the important role in the small $z$ region and are positive.

\begin{figure}[!ht]
\includegraphics[width=0.5\textwidth]{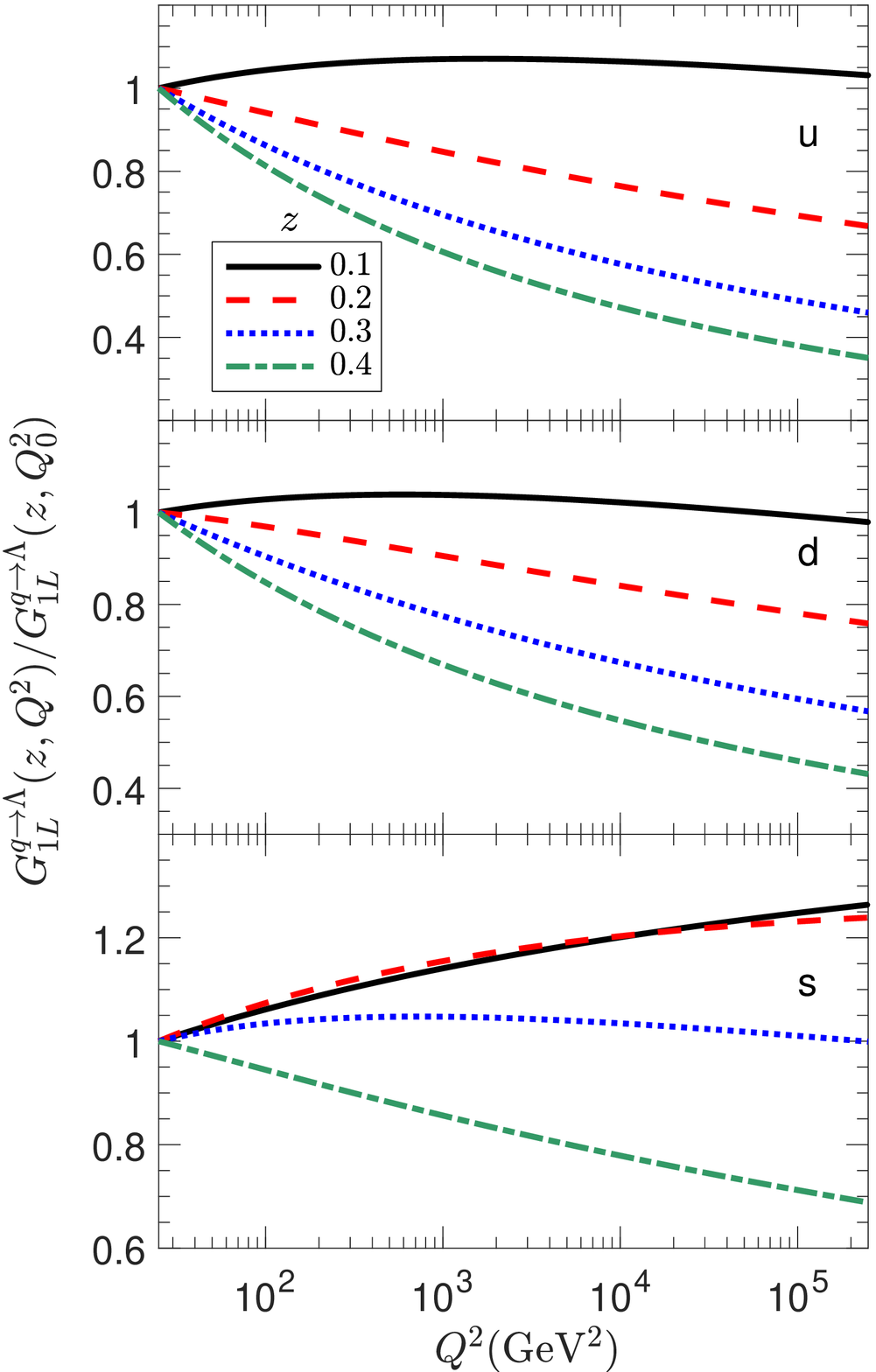}
\caption{(color online) QCD evolved $G_{1L}(z,Q^2)$ for $q \to \Lambda$ as function of $Q$ at different $z$ divided by the corresponding 
value at $Q_0=5$GeV.} 
\label{fig:evolved_G1Lq2lambda}
\end{figure}
\begin{figure}[!ht]
\includegraphics[width=0.5\textwidth]{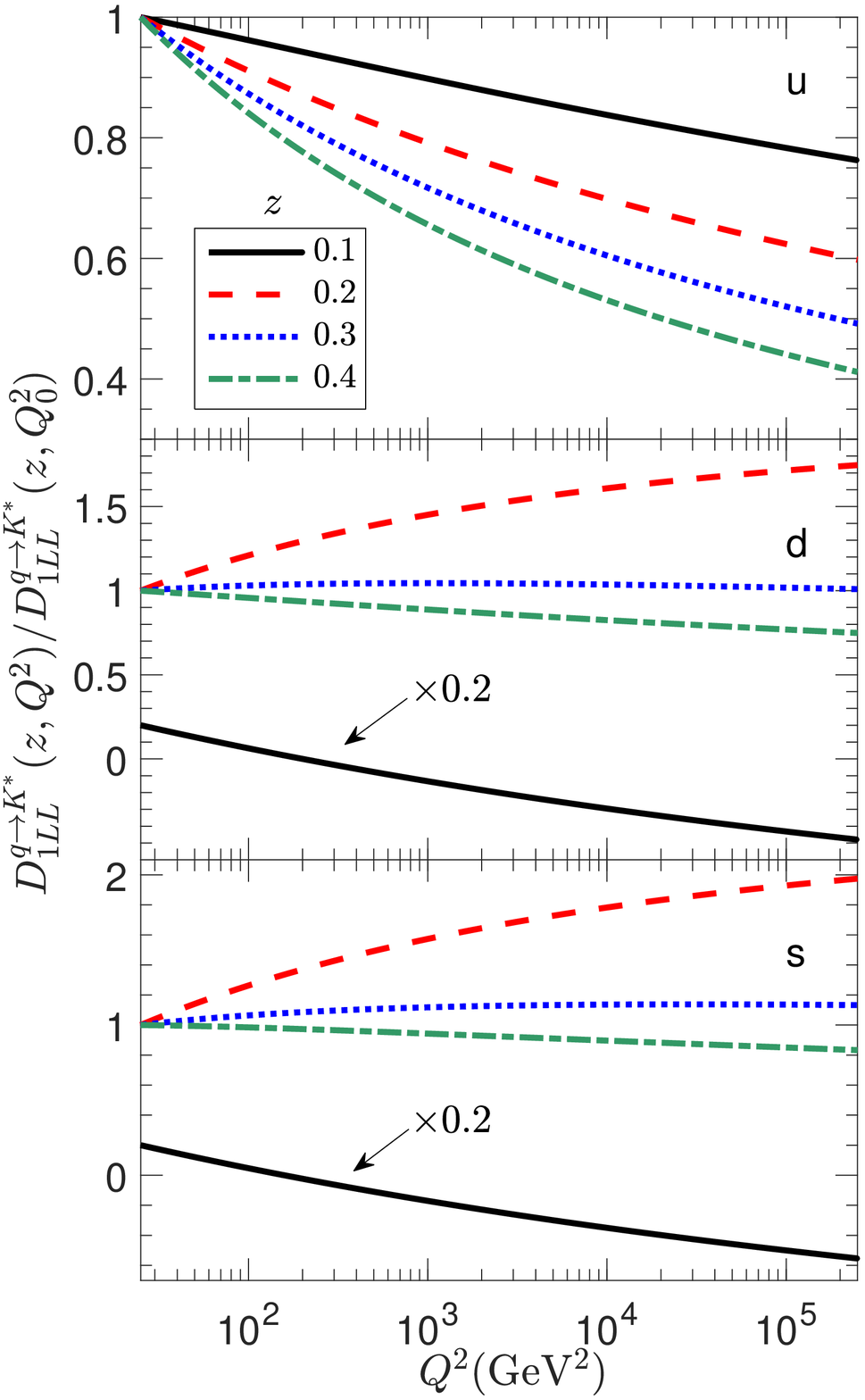}
\caption{QCD evolved $D_{1LL}(z,Q^2)$ for $q \to K^*$ as function of $Q$ at different $z$ divided by the corresponding value at $Q_0=5$GeV.} 
\label{fig:evolved_D1LLq2Kstar}
\end{figure}

From Fig.~\ref{fig:G1Lq2lambda}, we see that the peaks for $zG_{1L}(z,Q^2)$ shift towards smaller $z$ for larger $Q$. 
In the large $z$ region, $G_{1L}(z,Q^2)$ decreases with increasing $Q$ while for small $z$ it increases with increasing $Q$. 
This former is shown more obviously in Fig.~\ref{fig:evolved_G1Lq2lambda} where $G_{1L}^{q\to \Lambda}(z,Q^2)$ as a function of $Q$ at different values of $z$ is shown.
Similarly, in the large $z$ region, we see the same tendency for $D_{1LL}(z,Q^2)$ as a function of $Q$ from Fig.~\ref{fig:D1LLq2Kstar} and Fig.~\ref{fig:evolved_D1LLq2Kstar}, 
i.e. the magnitude of $D_{1LL}(z,Q^2)$ also decreases with increasing $Q$ for large $z$ values.
The relative rapid changes for the corresponding $D_{1LL}^{q\to K^*}(z,Q^2)$ at $z=0.1$ for $q=d$ or $s$ is due to the crossover with zero at $z\sim 0.1-0.2$. 
We see clearly that the magnitudes of these FFs do not change with $Q$ as drastically as $P_q$ does (see Fig~\ref{fig:Pq}). 
We therefore expect that the energy dependence of $P_{L\Lambda}$ should be dominated by that of $P_q$ 
and that of $\rho_{00}^{K^*}$ is smooth. 

\subsection{Energy dependence of $P_\Lambda$ and $\bar \rho_{00}^{K^*}$}\label{E-dep}
  
By inserting these results for FFs at different $Q$, we obtained $P_{L\Lambda}$ and $\bar \rho_{00}^{K^*}$ at different energies $\sqrt{s}=Q$. 
We plot the results in Figs.~\ref{fig:fit_Lambda} and \ref{fig:fit_SpinAlignment} respectively.  

\begin{figure}[!ht]
\includegraphics[width=0.5\textwidth]{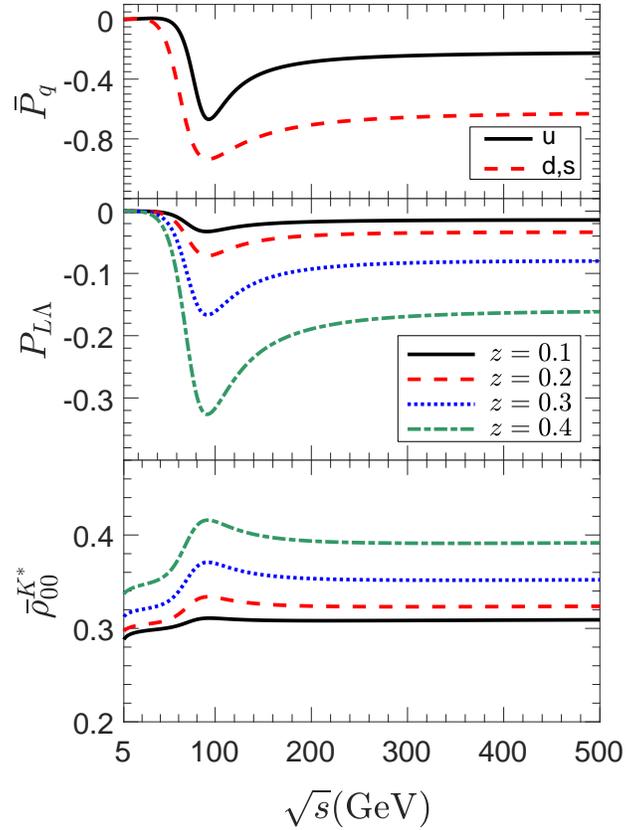}
\caption{(color online) Energy dependence of the longitudinal $\Lambda$ polarization in $e^+e^-$ annihilation.} 
\label{fig:energy_dep}
\end{figure}

From Figs.~\ref{fig:fit_Lambda} and \ref{fig:fit_SpinAlignment}, we see clearly that there is a strong energy dependence for $P_{L\Lambda}$, 
whereas that for $\bar \rho_{00}^{K^*}$ is quite weak. 
The former comes mainly from the energy dependence of $P_q$ while the latter comes mainly from QCD evolution of FFs.
To show this more explicitly, we plot $P_{L\Lambda}$ at a given $z$ as a function of $Q$ in the same figure as $\bar P_q$ in Fig.~\ref{fig:energy_dep}. 
For comparison, we also plot $\bar \rho_{00}^{K^*}$ in the 3rd panel of the same figure. 

From Fig.~\ref{fig:energy_dep}, we explicitly see that $P_{L\Lambda}$ behaves in very much the same way as $P_q$ as functions of $Q$.
This show clearly that the energy dependence of $P_{L\Lambda}$ is dominated by that of $P_q$. 
We see in particular that, just as $\bar P_q$, $\bar P_{L\Lambda}$ changes very fast with energy and goes to zero when $Q$ deviates from $Q=M_Z$ for $Q<M_Z$. 
This is because at smaller $Q$, electromagnetic interaction becomes dominant and weak interaction via exchange of $Z$-boson becomes negligible rapidly. 
Whereas at large $Q$, although smaller than that at the $Z$-pole, it is still sizable and becomes quite flat with increasing $Q$. 
The results show in particular that at BES or BELLE energies, $P_{L\Lambda}$ should be negligibly small. 
Furthermore, from the results presented in Sec.~\ref{sec:hp_energy} such as Eqs.~(\ref{eq:Lambda}-\ref{eq:Slty}), 
we see that there is no twist-3 contribution to $P_{L\Lambda}$ but there can be twist-3 contribution to the transverse components. 
Higher twist contributions to $P_{L\Lambda}$ come only from twist-4 or even higher twist~\cite{Wei:2013csa}.
This implies that at BES energies, the transverse components could even become larger than the longitudinal component 
for a given region of $\theta$ or $y$. 
 
 In contrast to $P_{L\Lambda}$, $\bar\rho_{00}$ changes with $Q$ quite weakly and remains sizable even at BES energies. 
 The relatively rapid change in the energy region around $M_Z$ comes from the influence of $W_q$. 
 This is a clear prediction that can be tested by future experiments~\cite{Abdesselam:2016nym}. 
 
 At the end of this section, we would like to emphasize once more the following. 
 Since the energy dependence of $P_{L\Lambda}$ is dominated by $P_q$, 
 the influence from other effects such as heavy flavor contribution, $u$ and $d$ flavor dependence, ``unfavored'' and gluon fragmentation etc 
 contribute only to the fine structure of the results shown in Fig.~\ref{fig:energy_dep}. 
 Lacking data and other related information, we simply neglected them at the initial scale in obtaining the results in Fig.~\ref{fig:energy_dep}. 
 However they are definitely worthwhile for experimental and theoretical studies in the future. 
 Furthermore, since they are addenda to the contribution from $P_q$ in the case of $P_{L\Lambda}$, 
 it might be more difficult to separate them from each other. 
 On the other hand, there is no contribution from $P_q$ for vector meson spin alignment $\rho_{00}$. 
 This means that such effects should play more important roles and manifest themselves more explicitly in different properties of $\rho_{00}$. 
 It could be much easier to study them in detail by studying $\rho_{00}$.
 In this sense, vector meson spin alignment could be a much better place to study different contributions in detail. 
 Furthermore, since it is independent of initial quark polarization, it is also forseeable that the effect of tensor polarization 
 determined by $\rho_{00}$ can also be studied in other high energy reactions in dependent of whether the initial hadron is polarized.  

\section{summary and outlook}

Using the longitudinal polarization $P_{L\Lambda}$ of $\Lambda$ hyperon and the spin alignment $\rho_{00}^{K^*}$ of $K^{*0}$ as representative examples, 
we demonstrate the two very different behaviors in energy dependences of hadron polarizations in $e^+e^-$ annihilations. 
The results show clearly that $P_{L\Lambda}$ has a very strong energy dependence due to its direct dependence 
on the initial longitudinal polarization $P_q$ of the quark $q$, 
while $\rho_{00}^{K^*}$ has a rather weak energy dependence since it is independent of $P_q$. 
The former is dominated by the energy dependence of $P_q$ while the latter comes mainly from the QCD evolutions of the FFs. 
We have presented the results at the leading twist with pQCD evolution at the leading order. 
In view that the measurements of both $P_{L\Lambda}$ and $\rho_{00}^{K^*}$ can in principle be easily carried out 
in experiments at BES or BELLE, we think that this provides a good place to test QCD evolutions of FFs and/or 
to check whether higher twist effects are important. 

\section*{Acknowledgements}
We thank S.Y. Wei for helpful discussions, C.Z. Yuan and X.P. Xu for communications. 
This work was supported in part by the National Natural Science Foundation of China
(Nos. 11675092 and 11375104),  the Major State Basic Research Development Program in China (No. 2014CB845406) 
and the CAS Center for Excellence in Particle Physics (CCEPP).

\end{document}